*Addendum to "$^{22}$Na production cross sections from the $^{19}$F (α,n) reaction"*

E. B. Norman, T. E. Chupp, K. T. Lesko, P. J. Grant, and G. L. Woodruff

In 1984, we reported $^{22}$Na production cross sections based on measurements of thick-target neutron yields from the $^{19}$F(α,n) reaction. Recently, there has interest in knowing the actual thick-target yields for applications such as assaying contents of UF$_6$ canisters using passive neutron counting techniques.[1,2] Here we report the results of our original thick-target neutron yield measurements over the energy range of E$_α$ = 3.5 − 10.0 MeV. As stated in our original paper and described in more detail in Ref. 3, we used several different techniques to determine our neutron detection efficiency to be 5.5 ± 0.3 % independent of neutron energy over the range we studied. One of these methods was to bombard a thick PbF$_2$ target with 10.0 MeV alpha particles. The neutron yield was measured during this bombardment and later the target was gamma counted to determine the $^{22}$Na activity produced. The gamma-ray detection efficiency was determined using two different calibrated $^{22}$Na sources. The neutron detection efficiency determined in this manner agreed to within ± 3.5% with those obtained by our other techniques.

In Table I we report our thick-target neutron yields as measured from a PbF$_2$ target. We then used the stopping powers of Andersen and Ziegler[4] to convert these yields into those expected from alpha bombardments of thick F$_2$ and UF$_6$ targets. Although we both used PbF$_2$ targets, our inferred results for F$_2$ are systematically higher than those of Bair and Gomez del Campo[5] by amounts that range from 54% to 35% as E$_α$ increases from 4 to 8 MeV. These differences could, at least in part, be due to differences in the stopping powers used. Note that Bair and Gomez del Campo did not report their measured PbF$_2$ yields or those expected from UF$_6$. However, as pointed out by Heaton *et al.*[6], our results are in good agreement with those of Feige *et al.*[7], as well as those reported by Sampson[8]. We also checked our technique by measuring thick-target (p,n) yields from Cd, Ta, and Pb at proton energies ranging from 5.5 to 9.5 MeV and comparing our results to those of Elwyn *et al.*[9]. Our data agree with those of Elwyn *et al.* to within ± 2.5%. We thus are confident of our results and believe that our thick-target neutron yield from UF$_6$ will be useful for assay applications.

**References:**

1. L.E. Smith, D.V. Jordan, A.C. Misner, E. Mace, C.R. Orton, *Enrichment Assay Methods for a UF6 Cylinder Verification Station*, IAEA-CN-184/125, (2011), IAEA, Vienna, Austria

2. H. O. Menlove, M. T. Swinhoe, and K. A. Miller, *A More Accurate and Penetrating Method to Measure the Enrichment and Mass of UF6 Storage Containers Using Passive Neutron Self-Interrogation,* (2010), LA-UR-10-03040, Los Alamos National Laboratory, Los Alamos, NM.

3. E. B. Norman *et al.,* Phys. Rev C **27**, 1728 (1983).

4. H. H. Andersen and J. F. Ziegler, *The Stopping Powers and Ranges of Ions in Matter* (Pergamon, New York, 1977), Vol. 4.

5. J. K. Bair and J. Gomez del Campo, Nucl. Sci. Eng. **71**, 18 (1979).

6. R. Heaton *et al.,* Nucl. Instrum. & Meth. A **276**, 529 (1989).

7. Y. Feige, B. G. Oltman, and J. Kastner, J. Geophys. Res. **73**, 3135 (1968).

8. T. E. Sampson, Nucl. Sci. Eng. **54**, 470 (1974).

9. A. J. Elwyn, A. Marinov, and J. P. Schiffer, Phys. Rev. **145**, 957 (1966).

**Table I.** Thick-target neutron yields from alpha-particle induced reactions in units of neutrons per $10^6$ alphas. Uncertainties are estimated to be $\pm 5\%$. Alpha energies are given in the laboratory frame. Note: 9.00E-2 = $9.00\times10^{-2}$.

| $E_\alpha$ (MeV) | Target | | |
|---|---|---|---|
| | **PbF$_2$** | **F$_2$** | **UF$_6$** |
| 3.50 | 9.00E-2 | 2.65E-1 | 1.54E-1 |
| 3.75 | 2.08E-1 | 6.17E-1 | 3.58E-1 |
| 4.00 | 4.11E-1 | 1.23E+0 | 7.11E-1 |
| 4.25 | 6.17E-1 | 1.84E+0 | 1.07E+0 |
| 4.50 | 1.02E+0 | 3.06E+0 | 1.77E+0 |
| 4.75 | 1.50E+0 | 4.52E+0 | 2.61E+0 |
| 5.00 | 1.96E+0 | 5.92E+0 | 3.41E+0 |
| 5.25 | 2.66E+0 | 8.07E+0 | 4.64E+0 |
| 5.50 | 3.45E+0 | 1.05E+1 | 6.03E+0 |
| 5.75 | 4.37E+0 | 1.33E+1 | 7.62E+0 |
| 6.00 | 5.31E+0 | 1.62E+1 | 9.28E+0 |
| 6.25 | 6.68E+0 | 2.05E+1 | 1.17E+1 |
| 6.50 | 8.00E+0 | 2.46E+1 | 1.40E+1 |
| 6.75 | 9.40E+0 | 2.89E+1 | 1.65E+1 |
| 7.00 | 1.08E+1 | 3.34E+1 | 1.90E+1 |
| 7.25 | 1.28E+1 | 3.96E+1 | 2.25E+1 |
| 7.50 | 1.46E+1 | 4.51E+1 | 2.56E+1 |
| 7.75 | 1.66E+1 | 5.12E+1 | 2.91E+1 |
| 8.00 | 1.86E+1 | 5.76E+1 | 3.27E+1 |
| 8.25 | 2.05E+1 | 6.36E+1 | 3.60E+1 |
| 8.50 | 2.23E+1 | 6.90E+1 | 3.90E+1 |
| 8.75 | 2.43E+1 | 7.56E+1 | 4.27E+1 |
| 9.00 | 2.65E+1 | 8.24E+1 | 4.66E+1 |
| 9.25 | 2.90E+1 | 9.01E+1 | 5.09E+1 |
| 9.50 | 3.12E+1 | 9.73E+1 | 5.49E+1 |
| 9.75 | 3.42E+1 | 1.06E+2 | 6.00E+1 |
| 10.0 | 3.64E+1 | 1.13E+2 | 6.39E+1 |